\documentclass{ws-procs975x65}

\def\beq{\begin{equation}}
\def\eeq{\end{equation}}

\begin{document}

\title{Pair Production, Vacuum Polarization and Anomaly in ${\rm (A)dS}$ and Charged Black Holes}
\author{Sang Pyo Kim}

\address{Department of Physics, Kunsan National University, Kunsan 54150, Korea\\
E-mail: sangkim@kunsan.ac.kr}

\begin{abstract}
We explore the connection between the distribution of particles spontaneously produced from an electric field or black hole and the vacuum persistence, twice the imaginary part of the one-loop effective action. Employing the reconstruction conjecture, we find the effective action for the Bose-Einstein or Fermi-Dirac distribution. The Schwinger effect in ${\rm AdS}_2$ is computed via the phase-integral method in the static coordinates. The Hawking radiation and Schwinger effect of a charged black hole is rederived and interpreted via the phase-integral. Finally, we discuss the relation between the vacuum persistence and the trace or gravitational anomalies.
\end{abstract}


\bodymatter


\section{Introduction}

The spontaneous production of  charged pairs in an electric field, the so-called Schwinger effect,\cite{schwinger} and the emission of all species of particles from a black hole, the so-called Hawking radiation,\cite{hawking} are the two most prominent nonperturbative quantum effects. Though the spectrum of radiation has been intensively investigated, the one-loop effective action has not been seriously studied yet. The in-out formalism by Schwinger and DeWitt provides a consistent and systematic theoretical framework to understand pair (particle) production and vacuum polarization.\cite{dewitt}

The gamma-function regularization,\cite{kim-lee-yoon08,kim-lee-yoon10,kim11} in particular, could explicitly yield the one-loop QED effective action in the proper-time integral\cite{heisenberg-euler,schwinger} from the Bogoliubov coefficient, which is expressed in terms of gamma functions. This method applies to the QED action in ${\rm (A)dS}_2$.\cite{cai-kim14} The near-horizon geometry of ${\rm AdS}_2 \times S^2$ of a near-extremal Reissner-Nordstr\"{o}m (RN) black hole sheds light on the charge emission from the near-extremal black hole.\cite{chen12,chen15,chen16} The Schwinger effect has been elaborated in ${\rm AdS}$ and charged black holes.\cite{kim14a,kim-lee-yoon15,kim15a,kim15b,kim15c,kim16a}

In this paper, we explore the connection between the particle distribution of spontaneously produced pairs and the vacuum persistence, twice the imaginary part of one-loop effective action, and construct the one-loop action for the Bose-Einstein or Fermi-Dirac distribution. It is noted that the vacuum persistence for the Bose-Einstein (Fermi-Dirac) distribution corresponds to that QED action for fermions (bosons). Employing the phase-integral method, we compute the Schwinger effect in the static coordinates of ${\rm AdS}$ and rederive the Hawking radiation of a non-extremal black hole and the Schwinger effect of a near-extremal black hole. Finally, we discuss the connection between the anomaly and the vacuum persistence.

\section{Particle Distribution, Vacuum Persistence and Polarization}

The particle or pair spontaneously created by an external electromagnetic field or black hole has the distribution
\begin{eqnarray}
N_{\lambda} = \frac{1}{e^{\beta \omega_{\lambda}} + \cos \theta},
\end{eqnarray}
where $\lambda$ denotes quantum numbers, and $\theta = 0$ for fermions, $\theta = \pi$ for bosons and $\theta = \pi/2$ for Boltzmann distribution. The parameter may be given by $\theta = (\pi/2)[1+ (-1)^{2 \sigma + i \tau_{\rm B}} ]$, where $\sigma (=0, 1/2)$ is the spin of particles and $\tau_{\rm B} = \infty$ for the Boltzmann distribution, otherwise $\tau_{\rm B} = 0$. In the in-out formalism, the vacuum persistence, twice of the imaginary part of the one-loop effective action, is related to the mean number of produced pairs of fermions and bosons as
\begin{eqnarray}
2 {\rm Im} {\cal L}^{(1)}_{\rm FD/BE} &=& (-1)^{2 \sigma} \sum_{\lambda}  \ln \bigl(1+ (-1)^{2 \sigma} N_{\lambda} \bigr) \nonumber\\
&=& (-1)^{2 \sigma + 1} \sum_{\lambda}  \ln \bigl(1+ (-1)^{2 \sigma + 1} e^{- \beta \omega_{\lambda}} \bigr). \label{FB vac per}
\end{eqnarray}
The summation implicitly includes the number of states as well as the sum over quantum numbers. The Boltzmann distribution does have the vacuum persistence
\begin{eqnarray}
2 {\rm Im} {\cal L}^{(1)}_{\rm B} = (-1)^{2 \sigma} \sum_{\lambda} \ln \bigl(1+ (-1)^{2 \sigma} e^{- \beta \omega_{\lambda}} \bigr). \label{B vac per}
\end{eqnarray}
In QED the vacuum persistence (\ref{B vac per}) is known as the Nikishov representation.\cite{nikishov}
Stephens noted that the vacuum persistence for the FD (BE) distribution corresponds to that of bosons (fermions) in QED.\cite{stephens} On the other hand, the the vacuum persistence for the Schwinger effect in a constant electric field could be expressed as an energy integral of the inverted spin statistics.\cite{hwang-kim} The Hawking radiation from black holes has either the Fermi-Dirac (FD) or Bose-Einstein (BE) distribution while the Schwinger effect due to a constant electric field in the Minkowski spacetime has the Boltzmann distribution.\cite{kim-hwang}

A conjecture of reconstruction of the one-loop effective action from the distribution of the produced mean number has recently proposed by the author and Schubert\cite{kim-schubert}
\begin{eqnarray} \label{cor}
{\cal L}^{(1)}_{\rm FD/BE} = (-1)^{2 \sigma + 1} \sum_{\lambda}  P \int_{0}^{\infty} \frac{ds}{s} e^{-\frac{\beta \omega_{\lambda}}{\pi}s} \Bigl( \frac{(\cos s)^{1-2 \sigma}}{\sin s} - g_{\frac{1}{2} - \sigma} (s) \Bigr),
\end{eqnarray}
where $P$ denotes the principal value and the function $g_{\sigma} (s) = 1/s - \cdots$ makes the effective action finite and corresponds to the renormalization of the vacuum energy and charge in QED. The Heisenberg-Euler and Schwinger QED action in a constant electric field is given by
\begin{eqnarray} \label{cor}
{\cal L}^{(1)}_{\rm B} = (-1)^{2 \sigma } \sum_{\lambda}  P \int_{0}^{\infty} \frac{ds}{s} e^{-\frac{\beta \omega_{\lambda}}{\pi}s} \Bigl( \frac{(\cos s)^{2 \sigma}}{\sin s} - g_{\sigma} (s) \Bigr).
\end{eqnarray}

\section{Pair Production in Static Coordinates of ${\rm (A)dS}_2$}

We study the Schwinger effect by a constant electric field in ${\rm (A)dS}_2$ space in the static coordinates
\begin{eqnarray}
ds^2 = - \bigl(1 - \Lambda r^2 \bigr) dt^2 + \frac{dr^2}{1 - \Lambda r^2}. \label{ads met}
\end{eqnarray}
The ${\rm AdS}_2$ is given by $\Lambda = - K^2 = R_{\rm AdS}/2$ and the ${\rm dS}_2$ is given by $\Lambda = H^2 = R_{\rm dS}/2$.
The constant electric field has the scalar potential $A_0 = -Er$, which leads to $dA_0 = E dt \wedge dr$ and $F_{01} = E = -F_{10}$. The equation for a scalar field with charge $q$ and mass $m$ in the constant electric field is given by
\begin{eqnarray}
\Bigl[- \frac{1}{1- \Lambda r^2}  \bigl(\partial_t + i qEr \bigr)^2 + \partial_r \bigl( 1 -  \Lambda r^2 \bigr) \partial_r  - m^2 \Bigr] \Phi (t, r) = 0.
\end{eqnarray}
We look for a solution of the Hamilton-Jacobi action $\Phi = e^{- i \omega t + i S_{\omega} (r)}$, which leads to the action
\begin{eqnarray}
S_{\omega} (r) = \int dr \frac{\sqrt{(\omega- eEr)^2 - m^2 (1 - \Lambda r^2)}}{1- \Lambda r^2}. \label{ads ac}
\end{eqnarray}

In Ref. \refcite{kim-page07} the leading term of the mean number of the Schwinger effect is given by the phase-integral $N_{\omega} = e^{i \oint dz S_{\omega} (z)}$ in the complex plane $z$ for $r$. The leading terms of particles produced with quantum number $\kappa$ in nonstatic coordinates have the form, $N_{\lambda} = \vert \sum_{J} e^{-i \oint_{C_{J}^{(1)}} dz \omega_{\lambda} (z)} \vert$, in the complex plane of time, where $C_{J}^{(1)}$ exhausts all possible independent contours of winding number one.\cite{kim13a,kim13b,kim14b} The higher winding numbers $C_{J}^{(n)}, (n \geq 2)$ will yield multi-pairs. In this paper we extend the geometric transition method for pair production to static coordinates of curved spacetimes
\begin{eqnarray}
N_{\lambda} = \Bigl\vert \sum_{J, n} e^{-i \oint_{C_{J}^{(n)}} dz S_{\lambda} (z)} \Bigr\vert. \label{geom tr}
\end{eqnarray}

The action (\ref{ads ac}) in ${\rm AdS}_2$ has two finite simple poles at $z = \pm i/K$ and another pole at $z = \infty$. The contour integral in Fig. 1 gives the mean number $N = e^{- {\cal S}_{\rm AdS}}$, where the instanton action is
\begin{eqnarray}
{\cal S}_{\rm AdS} = 2 \pi \Bigl( \frac{qE}{K^2}- \sqrt{\bigl(\frac{qE}{K^2} \bigr)^2 - \bigl(\frac{m}{K} \bigr)^2} \Bigr). \label{ads in}
\end{eqnarray}
The first term comes from the two simple poles $z = \pm i /K$ and the second term comes from another pole at $z = \infty$. In the absence of electric field the residues at $z = \pm i /K$ add up to zero and the residue from $z = \infty$ vanishes. The pure ${\rm AdS}$ is thus stable against the production of particles unless an electric field acts. The effective temperature for the Schwinger effect in ${\rm AdS}_2$ has recently been introduced by Cai and the author, which gives the physical interpretation
\begin{eqnarray}
N_{\rm AdS} = e^{- \frac{\bar{m}}{T_{\rm CK}}}, \quad \bar{m} = m \sqrt{1 - \frac{R_2}{8m^2}}, \quad T_{\rm CK} = T_{\rm U} + \sqrt{T_{\rm U}^2
+ \frac{R_2}{8 \pi^2} }, \label{ads tem}
\end{eqnarray}
with the Unruh temperature $T_{\rm U} = qE/(2 \pi \bar{m})$.
\begin{figure}[t]
\begin{center}
\includegraphics[width=2.5in]{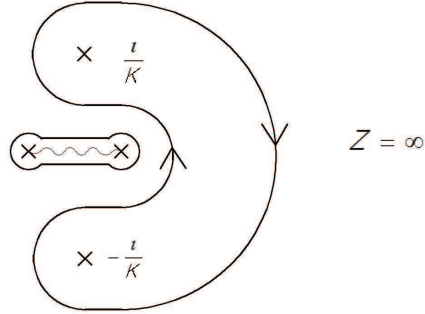}
\end{center}
\caption{The integral (\ref{ads ac}) of a charged scalar in a uniform electric field and ${\rm AdS}_2$ has a branch-cut in the complex plane $z$ and a pair of simple poles at $z= \pm i/K$ and another simple pole at $z= \infty$. The contour includes both poles at $z= \pm i/K$, and also receives a contribution from $z= \infty$.}\label{fig1}
\end{figure}

The action (\ref{ads ac}) in ${\rm dS}_2$  has two finite simple poles at $z = \pm 1/H$ and another simple pole at $z = \infty$. Applying the phase-integral (\ref{geom tr}), we find the leading term for the Schwinger effect
\begin{eqnarray}
{\cal S}_{\rm dS} = 2 \pi \Bigl( \sqrt{\bigl(\frac{qE}{H^2} \bigr)^2 + \bigl(\frac{m}{H} \bigr)^2} - \frac{qE}{H^2} \Bigr). \label{ds in}
\end{eqnarray}
The instanton has been found by the worldline instanton or the Bogoliubov transformation.\cite{garriga,kim-page08,cai-kim14,fishler}
The analytical continuation from $R_{\rm AdS} = -2K^2$ to $R_{\rm dS} = 2H^2$ holds only under the interaction of the electric field.
The Schwinger effect has also the thermal interpretation
\begin{eqnarray}
N = e^{- \frac{\bar{m}}{T_{\rm CK}}}, \quad \bar{m} = m \sqrt{1 - \frac{R_2}{8 m^2}}, \quad T_{\rm CK} = T_{\rm U} + \sqrt{T_{\rm U}^2 + T_{\rm GH}^2}, \label{ds tem}
\end{eqnarray}
where $T_{\rm GH} = H / 2 \pi$ is the Gibbons-Hawking temperature.

The Schwinger effect in the global coordinates of ${\rm (A)dS}_2$ differs that in the static coordinates.  In the global coordinates of ${\rm AdS}_2$
\begin{eqnarray}
ds^2 = - \cosh^2(Kx) dt^2 + dx^2
\end{eqnarray}
the Schwinger effect is given by\cite{kim-hwang-wang}
\begin{eqnarray}
N_{\rm AdS} = \Bigl(\frac{\sinh X}{\cosh Y} \Bigr)^2, \quad X = \pi \sqrt{\bigl(\frac{qE}{R_2/2} \bigr)^2 + \frac{m^2}{R_2/2} - \frac{1}{4}}, \quad Y = \pi \frac{qE}{\vert R_2/2 \vert}.
\end{eqnarray}
On the other hand, the Schwinger effect in the global coordinates of ${\rm dS}_2$ takes the form
\begin{eqnarray}
N_{\rm dS} = \Bigl(\frac{\cosh X}{\sinh Y} \Bigr)^2.
\end{eqnarray}
Note that the instanton action (\ref{ads in}) is ${\cal S}_{\rm AdS} = 2(Y-X)$ and that (\ref{ds in}) is ${\cal S}_{\rm dS} = 2(X - Y)$. In Ref. \refcite{kim15b} the phase-integral recovers the numerator and the leading term of the denominator. Note also that the leading term of the Schwinger effect is the same both in the static coordinates and in the global coordinates.

\section{Hawking Radiation and Schwinger Effect of Charged RN Black Holes}

The charged RN black hole with mass $M$ and charge $Q$ has the metric
\begin{eqnarray}
ds^2 = - f(r) dt^2 + \frac{dr^2}{f(r)} + r^2 d \Omega_2^2,
\end{eqnarray}
where
\begin{eqnarray}
f(r) = \frac{(r-r_{+})(r-r_{-})}{r^2}, \quad r_{\pm} = M \pm \sqrt{M^2 - Q^2}.
\end{eqnarray}
Note that particles are mostly emitted from the near-horizon region and the tunneling picture gives a physical interpretation of the Hawking radiation from there.\cite{parikh-wilczek} The near-horizon geometry of a non-extremal black hole has the geometry, ${\rm Rindler}_2 \times S^2$, while that of an extremal black hole has another geometry, ${\rm AdS}_2 \times S^2$. So the charge emission of the near-extremal black hole should be distinguished from that of the non-extremal black hole.

First, the metric for a non-extremal charged black hole can be written as the Rindler space\cite{kim07}
\begin{eqnarray}
ds^2 = - h (\rho) (\kappa \rho)^2 d t^2 +
\frac{1}{h(\rho) \bigl(1 + \frac{\rho^2}{r_{+}^2} \bigr)} d\rho^2 + r^2 (\rho) d\Omega_2^2,
\end{eqnarray}
where
\begin{eqnarray}
h(\rho) = \Bigl( \frac{2r_{+}}{r_{+} + r_{-} + \sqrt{(r_{+} - r_{-})^2 + (2 \kappa r_{+} \rho)^2}} \Bigr)^2.
\end{eqnarray}
Here, $\kappa$ is the surface gravity
\begin{eqnarray}
\kappa = \frac{r_{+}- r_{-}}{2 r_{+}^2}.
\end{eqnarray}
Contrary to the ${\rm (A)dS}_2$ in an electric field, the action (\ref{ads ac}) for a charged scalar has a single finite simple pole at $\rho=0$ with the residue $2 \pi i (\omega - qA_0) / \kappa$. Employing the phase-integral and counting all winding numbers, we may obtain the Hawking radiation
\begin{eqnarray}
N_{\omega} = \sum_{n=1}^{\infty} e^{- 2 \pi (\frac{\omega - qA_H}{\kappa}) n} = \frac{1}{e^{\beta (\omega - qA_H)} - 1},
\end{eqnarray}
where $\beta = 2 \pi/\kappa = 1/T_{\rm H}$ is the inverse Hawking temperature and $A_H = Q/r_{+}$ is the electric potential on the horizon. This implies that the Bose-Einstein distribution of produced particles is a consequence of a single finite simple pole. We may guess that the phase-integral for fermions should have an addition factor $e^{i \pi}$ to guarantee the Fermi-Dirac distribution. The gray factor may be related to the contribution from the infinity.

We may magnify the near-horizon region of a near-extremal black hole and write the geometry ${\rm AdS}_2 \times S^2$ as\cite{chen12}
\begin{eqnarray}
ds^2 = - \frac{\rho^2 - B^2}{Q^2} d \tau^2 + \frac{Q^2}{\rho^2 - B^2} d \rho^2 + Q^2 d \Omega_2^2,
\end{eqnarray}
where $\epsilon B = \sqrt{2Q(M-Q)}$ measures the deviation from the extremal black hole and the radial coordinate is stretched as $\rho = (r-Q)/\epsilon$ while the time coordinate is squeezed as $\tau = \epsilon t$. The squeezing of time does not affect the method due to the time-translational symmetry of the static nature. The action (\ref{ads ac}) for the $s$-wave gives the leading term\cite{chen12,kim15a,kim15b,kim16a}
\begin{eqnarray}
S_{\rm NBH} = 2 \pi \Bigl( qQ - \sqrt{(qQ)^2 - (mQ)^2 - \frac{1}{4(mQ)^2}} \Bigr). \label{ex bh}
\end{eqnarray}
Note that $E=1/Q$ and $R_H = -2/Q^2$ on the horizon, so $K=1/Q$ for ${\rm AdS}_2$. The action (\ref{ex bh}) is equivalent to the action (\ref{ads in}). The effective temperature for the Schwinger effect\cite{cai-kim14} gives a simple thermal interpretation
\begin{eqnarray}
N_{\rm S} = e^{- \frac{\bar{m}}{T_{\rm CK}}}.
\end{eqnarray}
where $T_{\rm U}$, $T_{\rm CK}$ and $\bar{m}$ are the Unruh temperature, the effective temperature and mass (\ref{ads tem}) with $R_2 = R_H$ on the horizon. The exact rates of charged boson and fermion emission from a near-extremal RN black hole have been explicitly studied in Refs. \refcite{chen12} and \refcite{chen15} and the charge emission from a near-extremal Kerr-Newman black hole has recently been found in Ref. \refcite{chen16}.
A thermal interpretation of the Schwinger emission of charged bosons and fermions has been advanced in Refs. \refcite{kim-lee-yoon15,kim15a,kim15b} and \refcite{kim16a}.

\section{Conclusion}

In the framework of quantum field theory we have critically reviewed the Hawking radiation as well as the Schwinger effect in ${\rm (A)dS}_2$ and a charged RN black hole. The quantum field theory should be able to properly include the effect of pair production in electromagnetic fields or black holes or de Sitter space since the vacuum there becomes unstable against the spontaneous emission of pairs (particles). The in-out formalism provides a consistent and comprehensive field theory at the one-loop or higher loops to study not only the vacuum polarization (real part of the effective action) but also the vacuum persistence (imaginary part). The one-loop effective action and the vacuum persistence exposed in this paper thus may shed light on understanding the Hawking radiation from a black hole and the Schwinger effect due an electric field. A charged black hole is a very interesting arena to probe the vacuum structure and to comprehend the intertwinement of quantum gravity and QED effects simultaneously.

The physical implications of the vacuum persistence have not been exploited in this paper. Ritus has proposed an interpretation in QED that the vacuum persistence (\ref{B vac per}) is the pressure of the fermion or boson gas.\cite{ritus} The fact that the vacuum persistence for fermion production is greater than that for boson production modulo the spin-multiplicity is consistent with the repulsive nature of fermions and the attractive nature of bosons. Another interesting aspect of nonperturbative quantum effect is anomalies. In QED the vacuum persistence is related to the imaginary part of the trace anomaly.\cite{dittrich-sieber}
The Hawking radiation is related to the trace anomaly.\cite{christensen-fulling}  The vacuum persistence for the Hawking radiation is the gravitational anomalies.\cite{kim-hwang} It would be an interesting open question to show whether the vacuum persistence in ${\rm (A)dS}$ and extremal black hole is related to the trace anomaly.

The Schwinger effect in ${\rm (A)dS}$ has been given a thermal interpretation with respect to the effective temperature associated with the acceleration of charges in the electric field.\cite{cai-kim14} It results from the solution of the field equation or the instanton action via the phase-integral method. The Unruh effect has been investigated in ${\rm (A)dS}$, in which the effective temperature is the geometric mean of the Unruh temperature and the Gibbons-Hawking temperature in ${\rm dS}$ or the scalar curvature in ${\rm AdS}$.\cite{narnhofer,deser-levin} The acceleration of charge by an electric field in ${\rm (A)dS}$ and charged black hole is determined by $qE/m$ and gives the Unruh temperature. One may be tempted to expect the effective temperature similar to the Unruh effect in ${\rm (A)dS}$. The effective temperature, however, involves an intrinsic nature as shown in eqs. (\ref{ads tem}) and (\ref{ds tem}). The worldline instanton in the embedding spacetime gives the same result as the phase-integral.\cite{samantray} In the Minkowski limit, the effective temperature is twice of the Unruh temperature for the accelerating charge in the electric field.

\section*{Acknowledgement}
The author thanks Rong-Gen Cai, Chiang-Mei Chen, Robert Mann, Don N. Page and Christian Schubert for useful discussions. He also thanks Won Kim for drawing the figure.
This work was supported in part by IBS (Institute for Basic Science) under IBS-R012-D1 and also by Basic Science Research Program through the National Research Foundation of Korea (NRF) funded by the Ministry of Education (NRF-2015R1D1A1A01060626).

\end{document}